\documentclass[runningheads]{llncs}
\usepackage{graphicx}
\usepackage{color}
\usepackage{colortbl, booktabs}
\usepackage{xcolor}
\usepackage{multirow}
\usepackage{subcaption}
\newif\ifdraft
 \drafttrue
\draftfalse
\ifdraft
   \newcommand{\todo}[1]{\textsf{\textbf{\textcolor{red}{[TODO: #1]}}}}
  \newcommand{\key}[1]{\textsf{{\textit{\textcolor{red}{#1:}}}}}
  \newcommand{\yuri}[1]{\textsf{\textcolor{blue}{\textbf{Yuri:} \textit{#1}}}}
  \newcommand{\takuya}[1]{\textsf{\textcolor{orange}{\textbf{Takuya:} \textit{#1}}}}
\else
  \newcommand{\todo}[1]{}
  \newcommand{\key}[1]{}
  \newcommand{\yuri}[1]{}
  \newcommand{\takuya}[1]{}
\fi

\begin{document}
%
% \title{Contribution Title\thanks{Supported by organization x.}}
\title{Stakeholder-in-the-Loop Fair Decisions: \\A Framework to Design Decision Support Systems in Public and Private Organizations}
\titlerunning{Stakeholder-in-the-Loop Fair Decisions}
\author{Yuri Nakao\orcidID{0000-0002-6813-9952} \and
Takuya Yokota}
\authorrunning{Y. Nakao and T. Yokota}
\institute{Fujitsu Limited, Kawasaki City, Japan\\
% \and Springer Heidelberg, Tiergartenstr. 17, 69121 Heidelberg, Germany
\email{nakao.yuri@fujitsu.com,  yokota-takuya@fujitsu.com}\\
}
\maketitle              % typeset the header of the contribution
\begin{abstract}
Due to the opacity of machine learning technology, there is a need for explainability and fairness in the decision support systems used in public or private organizations. Although the criteria for appropriate explanations and fair decisions change depending on the values of those who are affected by the decisions, there is a lack of discussion framework to consider the appropriate outputs for each stakeholder. 
In this paper, we propose a discussion framework that we call ``stakeholder-in-the-loop fair decisions.'' This is proposed to consider the requirements for appropriate explanations and fair decisions. We identified four stakeholders that need to be considered to design accountable decision support systems and discussed how to consider the appropriate outputs for each stakeholder by referring to our works. By clarifying the characteristics of specific stakeholders in each application domain and integrating the stakeholders' values into outputs that all stakeholders agree upon, decision support systems can be designed as systems that ensure accountable decision makings. 
\keywords{Machine Learning \and Fairness \and Explainability \and Decision Support Systems}
\end{abstract}

\section{Introduction}
Decision support systems using machine learning have expanded in private and public organizations and they have a great influence on our society.
The stakeholders of the decision support systems are diverse and many of them are non-experts in information technology, who have difficulty fully understanding the mechanism of  technology.
Despite that, each stakeholder needs to know how and why decisions that s/he is responsible for or that affect her or him, are made.
Hence, there is a requirement for interfaces that encourage smooth communication between the systems and their users based on the knowledge of human-computer interaction (HCI) techniques.

\subsection{Lack of Explainability}
To develop the interfaces, we have to overcome the social issues that machine learning provokes. 
Especially the issue of the lack of explainability and the issue of discriminatory bias are well-known.
The lack of explainability occurs due to the opacity of machine learning to its users or even to engineers \cite{clinciu2019survey}. 
Because, with complex methods such as deep learning, machine learning models are trained through a multilayered network, it is unable for human users to trace the whole process of the training. This leads to that no one can explain the reason for the output from the trained models.
This issue of lack of explainability is fatal to high-stakes social decisions such as medical diagnosis or prediction of recidivism.
To overcome this issue, the interpretability or explainability of machine learning has been explored. In the conventional research, there are approaches in which, after a machine learning model is trained, a different model to explain a local situation that its users want to know about is trained~\cite{Ribero_why_2016}, and that try to train machine learning models in understandable forms for human users~\cite{Lakkaraju_Interpretable_2016}.

In addition to the technology of interpretable and explainable machine learning, what kind of explanation is needed by the target users should be investigated.
The decision support systems used in private and public organizations have a variety of stakeholders. Hence, the systems' explanations should be changed so as to make themselves understandable for each stakeholder. For example, engineers in organizations can understand the statistical results explained quantitatively by the decision support systems. In contrast, non-expert users inside and outside the organizations do not understand such results. For non-expert users, the system should express the same results using natural languages.
How to express the appropriate explanation for each group of users is a major research topic that HCI community should investigate.

\subsection{Lack of Consensus on Fair Decisions}
On the other hand, the issue of discriminatory bias in machine learning processes has also been pointed out by academia and industry~\cite{Mehrabi_2021_Survey}. 
This issue occurs because, in the training data, there remain discriminatory biases in decisions in the past, such as gender bias or racial bias, and intersectional bias where such biases combine~\cite{Kobayashi_2022_OvO}. The machine learning models trained with biased data output discriminatory results. 
To remove the discriminatory bias from the training data or trained models, many technologies have been developed~\cite{Mehrabi_2021_Survey}.

One major concern about the fairness issue is the lack of consensus on the definition of fairness.
There are major two definitions of fairness: group fairness and individual fairness~\cite{Dwork_2012_Fairness}.
Group fairness is ensured when there is not any difference in the acceptance ratio or performance metrics between the protected group, i.e., discriminated groups such as a minority, colored people in Western countries, or people with disabilities, and the other group.
On the other hand, individual fairness is ensured when people who have the same ability or condition are treated the same.
It is known that these two definitions of fairness are not sometimes compatible. For example, when a person in a minority group is poorer than a person in a majority group, the person in a minority group will have less education and s/he will have a skill level that is lower than a person in a majority group. In this case, when we follow the definition of individual fairness, we should give preferential treatment to the person from a majority group. However, when we follow the definition of group fairness, we may be ought to give preferential treatment to a person from a minority group.

In addition to this kind of trade-off among the fairness definitions, there are different perceptions of fairness in different cultures.
For example, Kim and Leung~\cite{KIM_2007_Forming} clarified the cultural difference in the situations considered fair by people. According to them, employees in Japan and the U.S. tend to consider interactional fairness, which is the degree to which the people affected by a decision are treated with dignity and respect as fair. On the other hand, employees in China and Korea tend to consider distributive fairness, which is the fairness related to how rewards and costs are distributed across group members, as fair.
Hence, when we try to reach a consensus on fairness, we should consider to what cultural areas stakeholders belong.

In the context of the decision support systems used in public and private organizations, we should consider what are fair decisions for the stakeholders beyond the discussion of discriminatory bias.
According to Oxford Learners' Dictionary\footnote{https://www.oxfordlearnersdictionaries.com/}, the word `fair' means to be ``\textit{acceptable and appropriate in a particular situation}.''
The acceptable and appropriate conditions differ among the local situations of stakeholders.
For example, one group of stakeholders might think of the accurate decisions based on the historical data as fair, while another group might consider the decisions where discriminatory bias is removed as fair. 
Despite that, a social decision should ideally be what is considered fair by all stakeholders.
Hence, how to extract the condition stakeholders consider fair and how to integrate the perceived fair condition into the final decision is one of the important research topics of HCI.

\section{Our Focus}
In this paper, we focus on how to consider appropriate explanations and fair decisions output by the decision support systems used in public and private organizations. We classified the stakeholders who are related to the decision support systems into four groups: experts in organizations, direct recipients of decisions, indirect recipients of decisions, and regulators. Then, we discuss how to consider the explanations and decisions the stakeholders will agree upon. And, finally, we summarize the requirements for the explanations and decisions as the framework of stakeholder-in-the-loop fair decisions for further discussion. Through the framework, we contribute to the field of HCI to provide the foundation of discussion about how to design decision support systems in organizations.

\section{Explanation and Fairness for Each Stakeholder}
\subsection{Experts in Organizations}
Here, we consider the preferable explanation and fair results of the decision support systems to the stakeholders in the organizations.
First, there are two different types of stakeholders in the organization: information technology (IT) experts and domain experts.
IT experts are those who operate IT systems in organizations. Some of them manage the dataset and train machine learning models and adjust the models in their daily practice. Although IT experts sometimes exist in outsourced companies, here we consider the outsourced companies are a part of the organization for convenience.
On the other hand, domain experts are those whose roles are not related to IT, e.g., loan officers in a bank, who decide if a loan is approved for a customer or doctors in hospitals.
Although they do not usually operate decision support systems, they use the results from the systems to execute their roles. Because they usually have responsibilities for their role, they have to understand the results based on their own contexts.

Additionally, there are two different types of domain experts in organizations~\cite{Nakao_2019_Requirement}. First, there are domain experts who work on the deliverables that are used in different departments in the same organization. 
The deliverables become the resources on which the domain experts in other departments in the organization work based. 
Usually, there are various departments in one organization or company such as the customer service or citizen relation section, accounting section, and contract audit section. 
The final decisions for the outer customers or citizens are based on the integration of the decisions made by such various departments. 
To make the final decision accountable, domain experts in each department have the responsibility to make the accountable deliverables. 
Hence, they have to know the reason for their decisions. 
When decisions are made by an IT system, the reason for the decisions should be explained to the domain experts in a form that is reasonable and can be understood by the experts in other departments. 
The second type of domain experts is those who work for the outer customers or citizens directly, such as doctors in hospitals, and loan officers in banks.
Their works have effects on the people outside the organization. 
Works done by public organizations, such as local and national governments, affect the citizens' lives and works done by private companies affect the behavior of their customers.

It has been pointed out that different types of explanations are needed for different types of experts~\cite{arya_2019_one}.
For IT experts, statistical explanations are appropriate since IT experts have specialties in understanding statistical outputs. On the other hand, for domain experts, the systems' outputs should be explained in natural languages because the experts do not usually have the skills to understand the statistical outputs. 
At the same time, for the domain experts, the explanation should be made to meet the responsibility that the experts have. 
For example, of course, the explanation for the doctors and that for loan officers should be different.
Moreover, the appropriate explanations for the stakeholders inside the organizations have to be investigated to make HCI research match the daily practice of experts. 
The appropriate explanation for the experts who communicate with outer customers or citizens should be investigated because the experts' criteria are related to the accountability to the customers or citizens. 
Additionally, not only that, the accountability of the work done by the domain experts working for the experts in other departments inside the same organization should also be ensured because ensuring the accountability of the final decisions requires ensuring accountability in each phase of the decision process done in each department~\cite{Nakao_2019_Requirement}.

In addition to the different kinds of explanations required for different types of experts, different criteria for fair decisions are needed for the different types of experts. 
For example, an IT expert might consider accurate decisions meaning that the result from a decision support system matches historical data are fair. On the other hand, a domain expert, e.g., a loan officer, might consider the decisions that match the expert's intuition based on her/his experiences, e.g., workers in big companies tend to be approved in loan decisions, are fair.
Accordingly, the appropriate explanations and the criteria for fair decisions for the experts in organizations have to be investigated in each domain.

To investigate the similarity and differences inside a bank, we did research that explores a design space of user interfaces to support data scientists, i.e., IT experts, and loan officers, i.e., domain experts, to investigate the fairness of machine learning models~\cite{nakao2022TowardsResponsible}.
Using loan applications as an example, we held a series of workshops with loan officers and data scientists to elicit their requirements. 
As a result, for example, only data scientists need the information on sensitive attributes while loan officers need to feedback to data scientists on ``questionable'' attributes that should not be used for decision-making.
This result indicates that the data scientists consider decisions without discriminatory bias as fair while the loan officers consider the decisions that match their intuition as fair.
In the paper~\cite{nakao2022TowardsResponsible}, we proposed a test case of how to investigate the experts' viewpoints about explanations and fair decisions.

\subsection{Recipients of Decisions outside Organizations}
Next, there are two types of stakeholders outside the organizations who are affected by the decisions based on the outputs of decision support systems: direct recipients of decisions, and indirect recipients of decisions. 
Here, we discuss how to consider the appropriate explanation and fair decisions for them.

\subsubsection{Direct Recipient of Decisions}
First, there are direct recipients of decisions.
They receive and are influenced by the decisions made by domain experts in the organization directly.
The direct recipients are, for example, defendants in recidivism predictions, patients in medical diagnosis, recipients of investment in finance, and job candidates in job matching.
In the artificial intelligence ACT (AI ACT)~\cite{AI-ACT}, which is a draft of regulation to AI systems proposed by the European Commission (EC) in April 2021, AI systems except for the minimal risk AI have the obligation of transparency such as the obligation for chatbot systems to inform human users of s/he is interacting with AI systems.
Additionally, according to General Data Protection Regulation (GDPR)~\cite{GDPR}, data subjects, which are similar to the users of AI systems, ``\textit{should have enough relevant information about the envisaged use and consequences of the processing to ensure that any consent they provide represents an informed choice.}(Article 6(1)(a))''
Therefore, decision support systems in organizations have to be designed with a clear understanding of what explanations are necessary and sufficient for the direct recipient and what decisions are considered fair.

To clarify such explanations and criteria for fair decisions, now we discuss the types of direct recipients.
There are both recipients who have interests that are the same as and different from the experts in organizations have.
For example, in the medical context, patients, the direct recipients, might consider accurate decisions, i.e., diagnoses, in light of historical data as fair decisions. In this case, the patients consider accuracy, one of the performance metrics of machine learning, as the most important indicator for fair decisions.
This preference for performance metrics is the same as doctors, the domain experts in organizations who want to judge the remedies accurately.
On the other hand, there are also direct recipients who have different interests from the experts.
For example, in the context of recidivism predictions, judges, the domain expert in the court, want to make accurate decisions on the likelihood of recidivism. Hence, they will prefer to have the result from a decision support system whose accuracy is maximized.
On the contrary, defendants, the direct recipients of the decisions, do not want to be mistakenly judged as a person likely to re-offend. Therefore, they will prefer to  maximize the false-positive rate than the accuracy.
By clarifying if the interests of the direct recipients and the experts are the same or different, we can tell if the same or different explanations have to be expressed by the decision support systems and if the fair decisions are coherent or not between the experts and the direct recipients.

Moreover, when considering the direct recipients, we sometimes have to take global cultural diversity into account. When a private company such as a big bank has its branches globally, the decision support systems have effects on customers all over the world. In such cases, the concepts of fair decisions might be different based on cultural differences. For example, Geert Hofstede~\cite{hofstede2005cultures}, a social psychologist, developed a six-dimension model of national cultures based on global research and advocated that different countries have different tendencies in such dimensions as power distance, individualism, uncertainty avoidance, etc. If we follow the argument, for example, a person in a country which has a high individualism score might have a tendency to consider the denial of a loan application because of the arrest record of the customer's family member as unfair although a person who lives in a country whose individualism score is low might consider the loan decision is fair. Hence, when there are stakeholders globally, the difference in the cultural context should be taken into account.

To explore methods to investigate appropriate explanations and fair decisions for the direct recipients that exist globally, we did a research consisting of a series of workshops and crowdsourcing study~\cite{Nakao_Involving_2022}. 
Through workshops with end-users, we co-designed and implemented a prototype system that allowed end-users to see why predictions were made in a machine learning model of loan decisions, and then to change weights on features to debug fairness issues. We evaluated the use of this prototype system through a crowdsourcing study. 
To investigate the implications of diverse human values about fairness around the globe, we also explored how cultural dimensions might play a role in using this prototype. 
From this research, we found that cultural differences explained differences in assessing and improving fairness.
The cultural dimensions that seemed to matter most were Masculinity, Uncertainty Avoidance, and Indulgence in the Hofstedes' model~\cite{hofstede2005cultures}.
This research~\cite{Nakao_Involving_2022} is also a test case of how to investigate the global direct recipients' preference related to explanations and fair decisions.

\subsubsection{Indirect Recipients of Decisions}
Outside the organizations, there are also indirect recipients, who are indirectly affected by the decisions about the direct recipients. The indirect recipients are affected by decisions due to their relationships with the direct recipients. For example, in the domain of recidivism prediction, the members of the local community that a defendant belongs to are indirect recipients because they will be harmed if the decision, the prediction of recidivism, is wrong.
Similarly, insurance companies in the domain of medical diagnosis, recipients' business partners in loan decisions, or personnel placement agencies in job matching are the indirect recipients of decisions.

To consider appropriate explanations and fair decisions for the indirect recipients, we need to discuss the similarity and differences in the interests between the indirect recipients and other stakeholders. 
For example, in the medical context, an insurance company, which is an indirect recipient, does not want to pay the medical expenses for the erroneous diagnosis. Hence, when we consider the diagnosis that a patient has a disease as a positive instance, the company wants to minimize the false positive rate, which is the rate of patients diagnosed wrongly. 
On the other hand, in the domain of job matching, a personnel placement agency, an indirect recipient, wants to maximize its profits by receiving commissions for recruiting from the company where the candidate decides to be employed. 
Hence, they do not want a job candidate to be wrongly judged as an unqualified person.
In this case, when we consider the decision to hire a candidate as a positive instance, the personnel agency wants to minimize the false negative rate. This interest is the same as the job candidate, who does not either want to be judged wrongly as an unqualified candidate.

Based on the examples we discussed above, we can tell that the appropriate explanations and fair decisions for the indirect recipients change according to their interests. While that is the same as for other stakeholders, the obligation about what has to be considered also changes according to the domain of the decisions. When the public organization such as national or local governments possibly must not fail to consider the indirect recipients such as the local community in the case of recidivism predictions when generating explanations and fair decisions. This is because public organizations generally have to be accountable to the public. On the other hand, private companies, such as banks might not have to consider the indirect recipients when they make fair decisions due to the trade secret. Hence, for the decision support systems used in organizations, especially in public ones, there is the necessity to output that explains that their decisions are fair for the indirect recipients. The HCI community should explore how to generate the appropriate explanations for indirect recipients.

\subsection{Regulators outside the organizations}
The final stakeholder we consider is the external regulators outside the organizations. 
The regulators check if the organizations obey the laws, regulations, or constitutions regarding human rights. While they check the organizations that make social decisions entirely, some of the regulators are paying attention, especially to the algorithmic decision support systems used in the organization. In the famous example, ProPublica, which is known as a non-profit organization (NPO) for investigative journalism, pointed that the existence of racial bias in a recidivism scoring system used in the US court called COMPAS~\cite{Angwin_2016_MachineBias}. Although there are some criticisms of this report because the way of evaluating the discriminatory bias is not appropriate, the report was so influential that academia and industry started to focus on the fairness issues in the decision support systems.
Other than NPO, various countries have proposed their regulations on AI systems and have tried to control the decision support systems used in the organization to ensure the transparency, fairness, and accountability of social decisions to protect human rights~\cite{WhiteHouse_2022_Blueprint,Canada_2022_Responsible}.
Moreover, since GDPR and AI ACT, the regulations developed by EC have or will have effects globally, even if a country does not set any regulations on AI systems or data processing systems, companies, and organizations that try to operate globally are subject to control under those regulations. In this case, EC is a regulator.

Appropriate explanations and fair decisions for the regulators change depending on the laws or regulations on which the regulator is based.
An explanation is appropriate if the explanation provides the outer regulators with information enough to audit the process of decisions' lawfulness.
For fair decisions, what is needed by the regulators are fairness which means that there are not any discriminatory bias in the process or results of decisions. 
This kind of fairness can be ensured by using the conventional methods of fairness-aware machine learning~\cite{Mehrabi_2021_Survey}, which remove the bias based on sensitive attributes, such as race or gender, from training data or machine learning models.
However, since there are various types of fairness that should be ensured, such as group fairness or individual fairness~\cite{Dwork_2012_Fairness}, fairness in acceptance rate~\cite{barocas2016big} or in performance metrics~\cite{Hardt_2016_Equality}, and which sensitive attributes should be focused, the stakeholders should reach a consensus about the fairness that will be ensured.

To be responsible for the regulators' requirements, many companies declared their own AI Ethics guidelines~\cite{Microsoft_2022_Responsible,nakata2020initiatives}. With them, the companies try to show their attitude that they use AI technologies in ethical ways.
Moreover, several companies proposed methods to check ethical issues in the process of machine learning ~\cite{Madio_2020_Co-designing,IBM_2022_AIdesign,Nitta_2022_AIEthics}.
With these methods, people can come up with the potential ethical issues that can be evoked when machine learning and human users or society interacts.
Through these activities, the organizations which are mainly private companies try to ensure accountability for the outer regulators.

\section{Integration of Diverse Concept of Fairness}

Now, the authors are working on the research to design understandable and fair results of the decision support systems considering the preference of diverse stakeholders.
As we explained above, there are diverse stakeholders in decision support systems used in organizations. And different stakeholders have different skills and concepts of fair decisions. 
Since the decision support systems in the public and private organizations have a great impact on society, the difference in the preference for explanations and fairness should be intermediated. 
For the purpose of this intermediation, there are some previous studies that explore co-creation methods using workshops where multiple stakeholders meet in one place and discuss the preferable results of algorithmic systems~\cite{Lee_2019_WeBuildAI,Woodruff_2018_Qualitative}.
However, since making the diverse stakeholders get together in one place is difficult, and the number of people who participate in a workshop is limited, the workshop approach to extract preference for the machine learning models has limitations in terms of its scale.

Hence, we are now taking the crowdsourcing approach. Via crowdsourcing platforms such as Amazon Mechanical Turk\footnote{https://www.mturk.com/} and Prolific\footnote{https://www.prolific.co/}, we can access more diverse people than those who can meet in one place. We have explored what kind of situations are considered fair by diverse people via crowdsourcing using binary search method~\cite{nakao2022toward}, interactive systems with which users can evaluate the fairness of the decisions~\cite{Nakao_Involving_2022,Stumpf_2021_Design}, and choice of the preferable model seeing the performance metrics of machine learning models \cite{Yokota_Toward_2022}.
Although each method has its strength and drawbacks, we are now continuing to explore the best way to extract the diverse stakeholders' preferences for machine learning systems and integrate them in an agreeable form for all stakeholders.

\begin{figure}
\centering
\includegraphics[width=\textwidth]{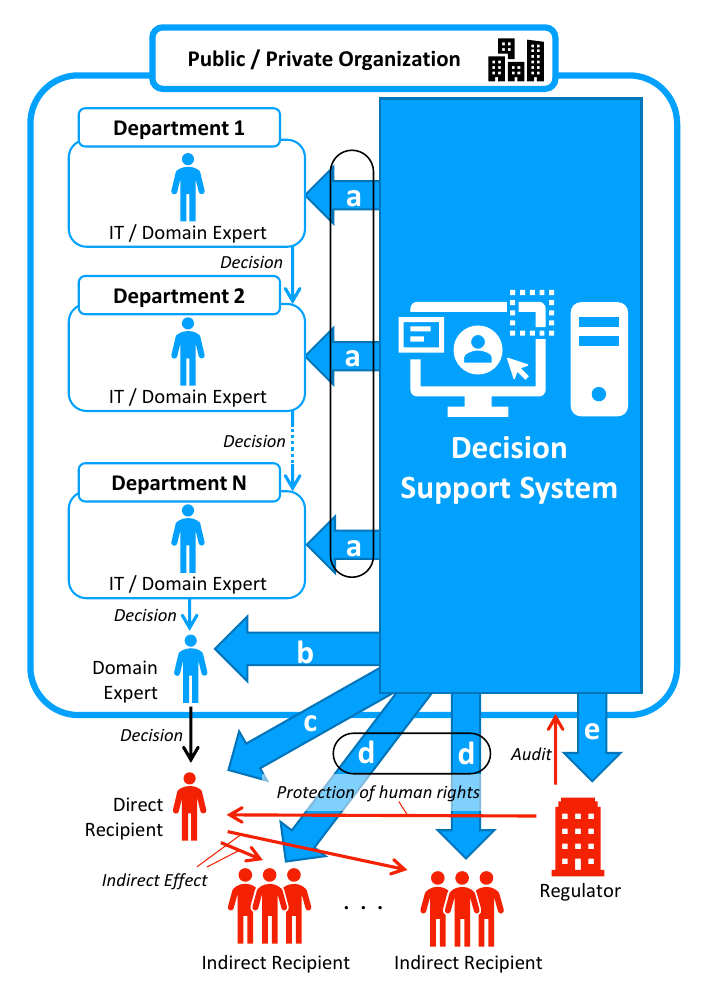}
\caption{Framework of the Stakeholder-in-the-loop fairness. This figure shows the relationships between stakeholders, and a decision support system in an organization that considers the stakeholders.} \label{fig1}
\end{figure}

\section{Framework of the Stakeholder-in-the-loop fairness}

Here, we summarize the requirements for the explanations and decisions we discussed above as the framework of stakeholder-in-the-loop fair decisions for further discussion.
In fig~\ref{fig1}, blue icons and frames indicate they are or belong to a public or private organization, and red icons indicate that they exist outside of the organization. The narrow arrows between icons indicate the relationship between stakeholders. And the bold arrows colored blue indicate the considerations for each stakeholder by the decision support system in the organization. The meanings of blue arrows as described as follows:

\begin{description}
\item[a] Considerations on the appropriate outputs for each expert. Each expert has her/his own skill and responsibility. Decision support systems' output should be accountable based on the skills and responsibilities. 
\item[b] Consideration of the domain expert who makes decisions for the outer customer or citizens. The expert is mainly the direct user of the decision support system. And the output for this domain expert should cover the values of all stakeholders.
\item[c] Consideration of the direct recipient of decisions. Decision support systems' outputs should be designed considering the direct recipients' situations and values. In some cases, the cultural background of the users should be cared about.
\item[d] Consideration of the indirect recipients of decisions. The indirect recipients are those who are affected by the direct recipients somehow. In some cases where the decisions are highly public, the indirect recipients, who are the part of citizens should be cared about.
\item[e] Consideration of regulators. The regulators try to protect the human right of the direct recipients and audit the activities of organizations based on laws and regulations. By developing the guidelines for the AI systems and using the tool to identify the potential issues, organizations can design the decision support systems as accountable systems that can respond to the regulators' requirements.
\end{description}

\section{Conclusion}
In this paper, we proposed a discussion framework that we name stakeholder-in-the-loop fair decisions. This framework is developed to consider the requirements for appropriate explanations and fair decisions obtained from the decision support systems in public and private organizations. We identified five stakeholders that need to be considered to design accountable decision support systems and discussed how to consider the appropriate outputs for each stakeholder by referring to our works. By clarifying the characteristics of specific stakeholders in each application domain and integrating the stakeholders' values into outputs that all stakeholders agree upon, decision support systems can be designed as systems that ensure accountable decision makings. To achieve accountability decision makings, authors will continue to work on this line of research.

\bibliographystyle{splncs04}
\bibliography{main}

\end{document}